**Sliding and superlubric moiré twisting ferroelectric transition in $HfO_2$**

Jie Sun[1*], Yiheng Shen[2*], Tengfei Cao[3], Li-Min Liu[4]

1. Graduate school of Engineering, The University of Tokyo, 7-3-1 Hongo, Bunkyo, Tokyo, 113-8656, Japan
2. Materials Genome Institute, Shanghai University, Shanghai 200444, China
3. Department of Materials Science and Engineering, Northwestern Polytechnical University, Xian 710072, China
4. School of Physics, Beihang University, Beijing 100191, China

Corresponding e-mail: sunjie113204@gmail.com    yihengshen@shu.edu.cn

## Abstract

Despite progress in $HfO_2$ thin-film ferroelectrics, issues like high coercive fields persist, and the dynamics of twisted ferroelectricity remain largely unexplored. Here, we explore how sliding and twisting in bilayer $HfO_2$ enables low barrier switching. Among 144 sliding configurations, two exhibit strong in-plane polarization (2360 pC/m) with a low switching barrier of 9.57 meV/f.u. Twisting generates polar textures associated with moiré patterns, which drive ferroelectricity via a soft zone-center mode, as revealed by machine-learning-assisted first-principles calculations. The in-plane (out-of-plane) polarization values for $HfO_2$ at twist angles of 21.79°, 27.80°, and 46.83° are 430 (5.82), 367 (2.20), and 1057 (0.03) pC/m, respectively. For 21.79° and 27.80° twisting, switching barriers drop to 1.74 and 0.18 meV/f.u., indicating superlubric-like transitions. Notably, the 46.83° twisted bilayer shows an almost barrier-free polar evolution (0.03 meV /f.u.), attributed to sharply enhanced zone-center phonon linewidths. Our findings establish a moiré-engineered switching route for 2D ferroelectrics.

## Introduction

Ferroelectric materials have been widely explored for energy-efficient nanoelectronics [1,2], including logic devices like ferroelectric field-effect transistors [3] and negative capacitance devices [4,5], as well as nonvolatile memory elements such as memristors [6], and ferroelectric tunnel junctions [7]. Among them, $HfO_2$-based ferroelectrics stand out for their robust polarization down to the unit-cell level [8] and exceptional compatibility with CMOS processes [9,10], in contrast to conventional perovskite ferroelectrics whose performance degrades at reduced thickness [11-13]. However, despite these advantages, several challenges remain. The relatively high coercive field hampers efficient energy switching, while charge trapping and defect accumulation driven by high external field contribute to ferroelectric fatigue, raising concerns about the long-term endurance and reliability of these devices [14]. Moreover, current research remains largely confined to epitaxial or polycrystalline thin films, where ferroelectric properties are inherently limited by structural rigidity [15], substrate-induced strain [16], and symmetry constraints [17]. These limitations underscore the need for new design strategies that can transcend the architecture of conventional thin films.

Initially developed in van der Waals heterostructures [18-20], ultra-low energy barrier interlayer sliding, and twisting were introduced as the key approaches to break inversion symmetry and engineer ferroelectricity [21,22]. Recent breakthroughs in the synthesis of free-standing oxide membranes [23] and the fabrication of bilayer perovskite membranes heterostructures, either through sliding [24,25] or twisting with moiré superlattices [26], have opened new degrees of freedom for exploring ferroelectric phenomena beyond the constraints of substrates. Previous work on twisted bilayer freestanding oxide membranes have revealed a range of novel phenomena, including moiré-induced interfacial ferroelectric polar vortices [27-29], magnetism modulation, flat electronic bands, topological Lieb lattices, as well as the topological superconductivity [27,30]. These findings highlight the potential of twist-enabled interfacial engineering in oxide systems to generate collective states and emergent functionalities for next-generation electronic and memory devices. Despite these advances, interfacial sliding and moiré engineering in freestanding $HfO_2$ systems remain unexplored.

In this work, to address the high computational cost of capturing dynamic ferroelectric transitions in twisted systems, we combine first-principles calculations with machine learning techniques to investigate, for the first time, the modulation of interfacial ferroelectricity in bilayer $HfO_2$ under both sliding and twisting configurations. We demonstrate that interlayer sliding and twisting in bilayer $HfO_2$ enables ultra-low barrier ferroelectric switching. Two stable sliding configurations exhibit strong in-plane polarization with a low barrier of 9.57 meV/f.u. Twisting generates polar texture associated with moiré pattern that activates soft optical phonons, reducing switching barriers to as low as 0.18 meV/f.u., smaller than that in the superlubric ferroelectric transition in $MoS_2$ [31]. These results reveal a moiré-engineered pathway to overcome fatigue and coercive field limitations in nanoscale ferroelectrics.

## Results

### Geometric structure and phonon dispersion

The phase diagram of bulk $HfO_2$ is formed by the polymorphs, in which the metastable $Pca2_1$ phase being recognized as the origin of its ferroelectricity. Recent experiments have demonstrated that most $Pca2_1$-structured $HfO_2$ can be grown along the [1,1,1] direction [10,13], forming an ultrathin film with a thickness of only two atomic layers and the freestanding $HfO_2$ can be maintained down to 1 nm or thinner [8,32]. Inspired by this, we theoretically cleaved the $Pca2_1$ phase along this direction to obtain a monolayer $HfO_2$ structure, which we refer to as the 1T phase, as illustrated in Fig. 1(a). The optimized 1T monolayer $HfO_2$ adopts a $p\bar{3}m1$ symmetry with lattice constants $a = b = 3.24$ Å, using the density functional theory (DFT) calculations. The detailed settings are seen in the Methods. Due to its centrosymmetric nature, it lacks spontaneous polarization, meaning it does not exhibit intrinsic ferroelectricity. We then investigated the impact of stacking configuration on polarization behavior in bilayer $HfO_2$. The commonly seen stacking pattern of 1T phase is the AA and AB stacking configurations, as shown in Fig. 1(b) and (c). The optimized lattice constants for AA and AB stacking are 3.23 and 3.24 Å, and the corresponding interlayer distance (defined by the vertical distance between O atoms in the two layers) are 2.76 and 2.40 Å, respectively. The AA stacking is less stable than the AB stacking by 37.60 meV /f.u. When $HfO_2$ is stacked in the AA or AB configuration, the structure maintains inversion symmetry, preventing the emergence of polarization. By applying lateral mirror operation to one layer in AA stacking, the inversion symmetry is broken, enabling ferroelectric polarization upon sliding. This distinct stacking configuration is referred to as AA' stacking, with side views as depicted in Fig. 1(d). The optimized lattice constant of the bilayer AA' $HfO_2$ is 3.23 Å, exhibiting a symmetry of $p\bar{6}m2$. Within the harmonic approximation, the phonon dispersion of AA' $HfO_2$ exhibits some soft modes, as seen in Fig. 1(e), which can be hardened by the temperature. To verify the dynamic stability of the AA' stacking phase, the self-consistent phonon theory is employed to renormalize the phonons while accounting for temperature effects [33]. The resulting phonon dispersion at room temperature confirms the stability of the AA' stacking, which is calculated with the help of self-consistent phonon (SCPH) theory [33], as implemented in HiPhive package [34].

The projected electronic band structure, presented in Fig. S1(a) in the Supplementary Information (SI), shows that monolayer $HfO_2$ exhibits a wide bandgap of 4.77 eV. The valence band is mainly contributed by *p* orbitals of O atoms while the *d* orbitals of Hf atoms dominate the conducting bands according to the projected density of states (PDOS). Fig. S1(b) shows the band structure of the bilayer $HfO_2$ with AA' stacking, with a band gap of 4.62 eV, slightly smaller than that of the monolayer $HfO_2$, which can be attributed to the band smearing over the duplication of analogous 2D lattices. In general, the similar band dispersion of monolayer and bilayer $HfO_2$ indicates the weak interlayer interaction between the layers, in accordance with the interlayer distance of 3.15 Å.

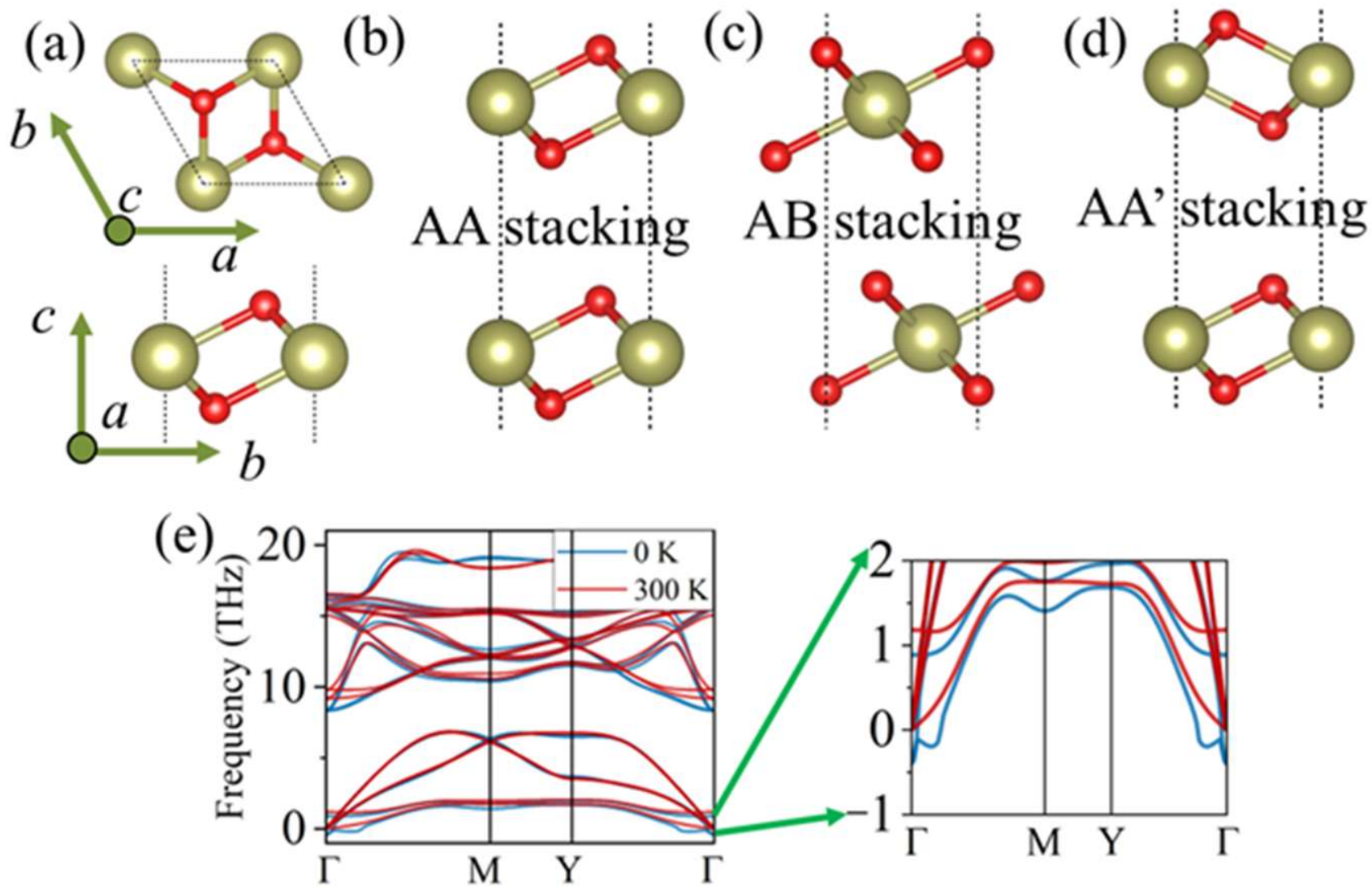

**Fig. 1. Structural, vibrational, and electronic properties of monolayer and bilayer $HfO_2$.** (**a**) Top and side views of monolayer $HfO_2$ (1T phase). Side views of bilayer $HfO_2$ with (**b**) AA stacking, (**c**) AB stacking and (**d**) AA' stacking. (**e**) Phonon dispersions of bilayer $HfO_2$ (AA') at 0 K and 300 K, with the inset highlighting the frequency range from -1 to 2 THz.

**Sliding ferroelectrics**

Starting from non-twisted AA' bilayer $HfO_2$, we generated 144 unique stacking patterns by shifting one layer relative to the other within a 12×12 displacement grid along the lattice vectors **a** and **b** (defined in Fig. 1(a)). Each stacking configuration is represented by a displacement vector $\mathbf{r}=m\mathbf{a}+n\mathbf{b}$, where $m$, $n$ $\in$ $\{x/12 \mid x = 0, 1, 2, \ldots, 11\}$ correspond to discrete displacement steps. After geometric optimizations using DFT, the energy contour of the stacking patterns is plotted in Fig. 2(a). Among all stacking configurations, we found that the structures labeled (1/3, 2/3) and (2/3, 1/3) exhibit the lowest energy, indicating their potential stability as ground states. These configurations, denoted as $AA_1$ and $A_1A$, adopt a *P3m1* symmetry, as shown in Fig. 2(b). The stable phonon dispersions within the harmonic approximation indicate the dynamic stability of the two configurations, shown in Fig. 2(c) and (d). The corresponding electronic band structures, presented in Fig. S2, reveal a common band gap of 4.68 eV, which is close to that of the AA' stacking one. Further calculations of spontaneous polarization using the Berry phase method [35,36] confirm that the $AA_1$ and $A_1A$ configurations exhibit an out-of-plane polarization of ±1.95 pC/m and in-plane polarization of ±2360 pC/m. The unit used here is commonly used for 2D ferroelectric systems [25,37-39]. The out-of-plane polarization is slightly smaller than those in bilayer *h*-BN (2 pC/m) and $MoS_2$ (5 pC/m) [25,40], but both of which exhibit zero in-plane polarization. This highlights the emergence of pronounced sliding ferroelectricity in bilayer $HfO_2$.

Another crucial factor in evaluating ferroelectric performance is the energy barrier of polarization switching. Our Nudged Elastic Band (NEB) [41] calculations show that the switching barrier from $AA_1$ to $A_1A$ is only 9.57 meV/f.u., indicating that polarization reversal is highly accessible. The evolution of energy along the switching pathway is illustrated in Fig. 2(e). The displacement index in Fig. 2(e) denotes the sliding steps along the NEB-calculated transition paths: AA' → $AA_1$ (indices 0–4), $AA_1$ → $A_1A$ (indices 4–8), and $A_1A$ → AA' (indices 8–12). The phonon dispersion of the intermediate state corresponding to displacement index 6 is shown in Fig. S3, whose soft optical mode drives the ferroelectric transition, as detailed in the SI. The corresponding geometric structures along the polarized path from $A_1A$ stacking (sliding index 4) to $AA_1$ (sliding index 8) are shown in Fig. S4. Interestingly, while the AA' stacking exhibits zero polarization, its higher energy suggests that it is not preferred as an intermediate state in the displacement pathway of bilayer $HfO_2$.

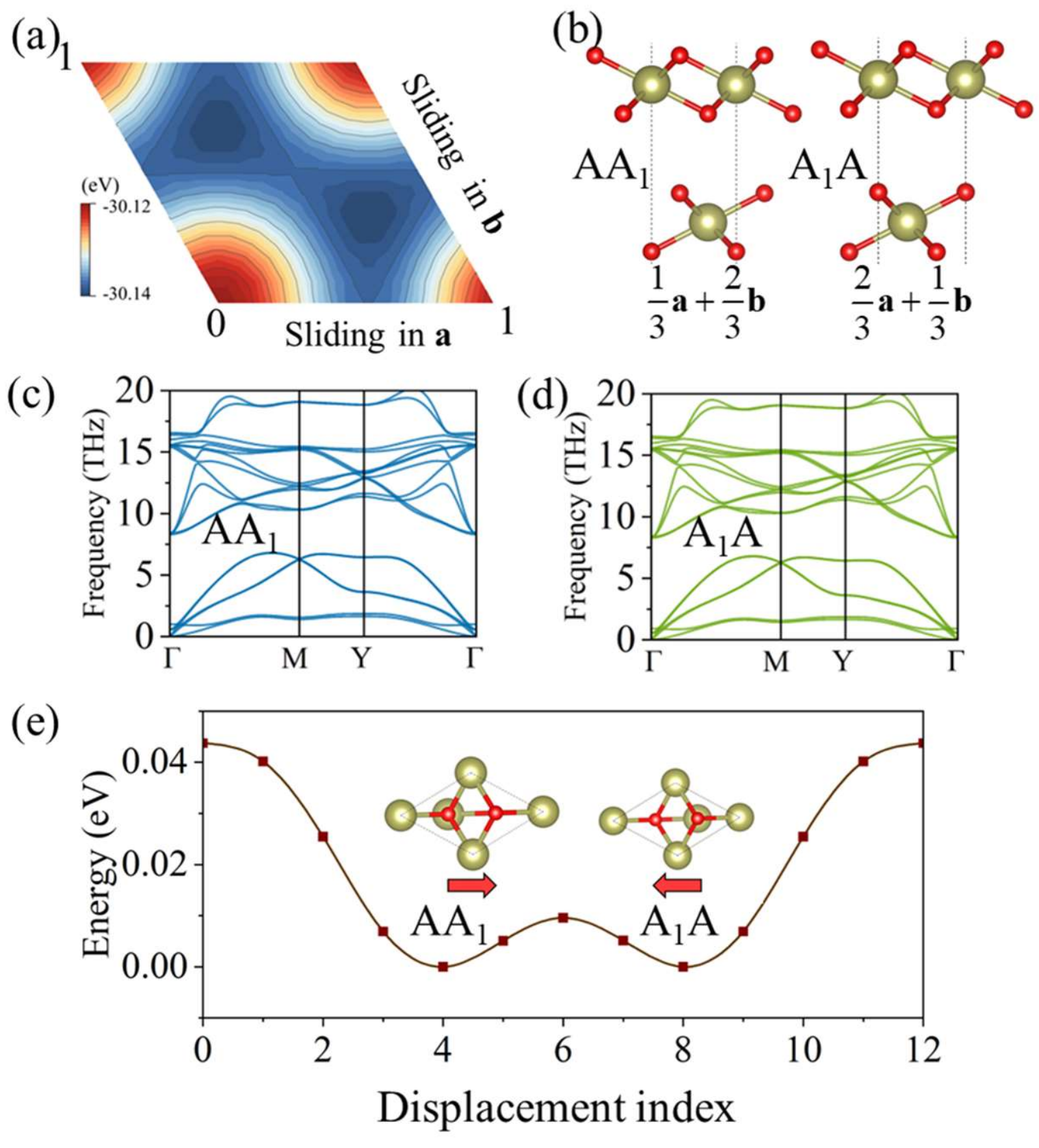

**Fig. 2. Sliding energy landscape and phonon properties of bilayer $HfO_2$.** (**a**) Energy contour (eV/f.u.) plot depicting the sliding configurations of a bilayer $HfO_2$ as a function of the bottom layer's displacement relative to the up layer. (**b**) Side view of the two lowest energy configurations, with indices (1/3, 2/3) and (2/3, 1/3), referred as $AA_1$ and $A_1A$, respectively. Phonon dispersions for the (**c**) $AA_1$ and (**d**) $A_1A$ configurations, respectively. (**e**)

Energy variations of different sliding patterns from $AA_1$ to $A_1A$ stacking, with the inset showing the top views of the two configurations.

**Twisting ferroelectrics**

We then introduced the twisting strategy in the bilayer AA' $HfO_2$ to investigate potential twisted ferroelectricity. Using the *Cell Match* code [42] to identify a common lattice by expanding the primitive cells of both layers, six twisted supercell structures with twisting angles of 13.17°, 21.79°, 27.80°, 32.20°, 38.21°, and 46.83° are selected within the 0°-60° range. Twisting by 13.174° and 46.826°, 21.79° and 38.21°, as well as 27.80° and 32.20° leads to identical unit cell expansions, as illustrated in Fig. S5. Here, we focus on structures with twisting angles of 21.79°, 27.80°, and 46.83° for further study, with their top views presented in Fig. 3(a-c). Interestingly, the twisted structures constructed here resemble those of $NiI_2$ due to their shared symmetry of *P3m1* [43]. The optimized lattice constants for the twisted 21.79°, 27.80°, and 46.83° are 8.55, 11.65 and 14.08 Å, respectively. The corresponding interlayer distances are 2.81, 2.82 and 2.76 Å, respectively, which are all decreased as compared to that of the untwisted structures due to stronger interlayer interactions.

Inspired by the observed chiral vortex pattern in the twisted bilayer $BaTiO_3$ [27], we investigated the potential polar texture associated with moiré pattern in bilayer $HfO_2$. For this purpose, we analyzed the in-plane displacement vector of O atoms ( $\Delta\mathbf{r}$ ), defined by $\mathbf{r}_{relaxed} - \mathbf{r}_{initial}$ , where $\mathbf{r}_{initial}$ and $\mathbf{r}_{relaxed}$ represent the positions of atoms before and after structural optimization. The displacement vectors of O atoms for bilayer $HfO_2$ with twisting angles of 21.79°, 27.80°, and 46.83° are shown in Fig. 3(d-f), revealing the emergence of polar textures across all twist angles. Notably, the polar textures in the upper (red color) and lower layers (blue color) exhibit almost opposite rotational orientations, resembling those in twisted bilayer $BaTiO_3$ [27]. The out-of-plane displacement of atoms O also follows an antiparallel alignment between the upper and lower layers, as illustrated in Fig. S6.

Fig. 3(g-i) presents the electronic band structures of bilayer $HfO_2$ with twisting angles of 21.79°, 27.80°, and 46.83°. The corresponding band gaps for 21.79° and 27.80° are both approximately 4.79 eV, while for 46.83°, the band gap decreases to 4.65 eV. In all three twisted structures, the valence bands become noticeably flatter compared to that of the untwisted structure shown in Fig. S(1). To further analyze this effect, we zoom in on the valence bands within a narrower energy range from -0.5 eV to 0.0 eV. The bandwidth initially decreases and then increases with an increasing twist angle, reducing from 73 meV to 25 meV before rising to 53 meV. The twisted $HfO_2$ at 27.80° exhibits the smallest bandwidth, approaching the quasi-flat bands observed in bilayer TMDs [44]. The bandwidth is defined as the difference between the maximum and minimum energy values of the valence band. The maximum (minimum) energy values in the valence band for the twisted 21.79°, 27.80° and 46.83° structures are -0.134 eV (-0.242 eV), -0.07 eV (-0.107 eV), -0.096 eV (-0.149 eV). Therefore, the corresponding bandwidth are 108 meV, 37 meV, 53 meV, respectively.

Previous work defined the bandwidth of 25-150 meV as the quasi-flat band [45], our calculated bandwidths are in this range, suggesting the quasi-flat bands in the twisted structures.

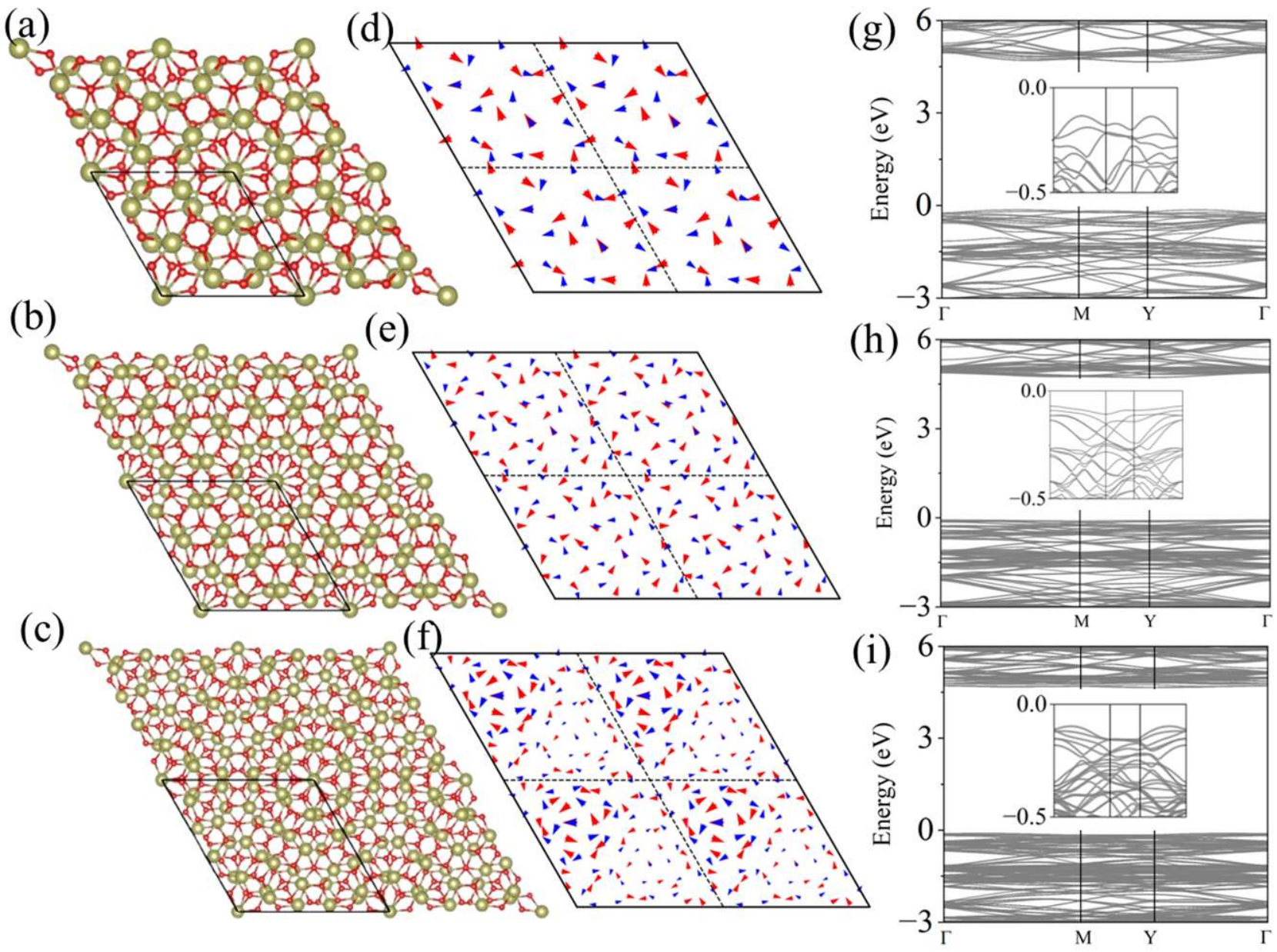

**Fig. 3. Moiré patterns, polar textures, and electronic structures of twisted bilayer $HfO_2$.** Top views of twisted bilayer $HfO_2$ (2×2×1 supercell) with twisting angles of (**a**) 21.79°, (**b**) 27.80°, and (**c**) 46.83°. (**d-f**) Corresponding polar textures associated with moiré patterns, where the red and blue colors represent the oxygen displacement in upper and lower layers, respectively. The magnitude of the displacement is enlarged by three times for clearer view. (**g-i**) Band structures of twisted bilayer $HfO_2$ for the corresponding twisting angles, with insets highlighting the energy range from -0.5 to 0.0 eV.

Considering the common presence of strain gradients in twisted systems [46-48], we started by introducing strain gradient (in the form of equation 1 in Methods) to explore the polar texture in twisted $HfO_2$. However, it induced dynamical instability evidenced by imaginary phonon modes across the first Brillouin zone, as shown in the 21.78° configuration in Fig. S7. Therefore, strain gradient here serves only as a perturbation, necessitating further structural optimization. We used machine learning potential, more specifically, Neuroevolution potential (NEP) [49], to optimize the twisted structures, and found that the inversed strain settings led to optimized configurations (See Fig. S8) with opposite polarization for all twist angles (See Fig. 4). The in-plane (out-of-plane) polarization for twisted bilayer $HfO_2$ with twist angles of 21.79°, 27.80°, and 46.83°, are 430 (5.82), 367 (2.20), and 1057 (0.03) pC/m, respectively. The details of the NEP potential training process can be found in Methods and Fig. S9.

To explore the origin of the ferroelectricity in the twisted structures, the track on the change of phonon frequency is an important perspective during the phase transition. Here, we used the above-mentioned NEP potential, to calculate the atomic forces in the dynamical matrix of twisted $HfO_2$. The phonon dispersions of the three twisted structures without strain perturbation are shown in Fig. S10, with the phonon frequency range magnified from -0.2 THz to 0.2 THz. For twisted 21.79° and 27.80° $HfO_2$, a zone-center optical soft mode is observed, indicating the origin of the phase transition. After strain perturbation, the soft mode frequency can be lifted above 0 THz, as shown in Fig. 4 and Fig. S11. Unlike the 21.79° and 27.80° twisted structures, the 46.83° system exhibits no imaginary frequencies. Interestingly, we also observe that under sufficiently large strain perturbation, twisted bilayer $HfO_2$ can transform into dynamically stable amorphous-like structure, as shown in Fig. S12.

We then performed NEB calculations for the three twisted structures, as shown in Fig. 4. The energy of the structure is calculated based on the NEP potential based on *cg* method in GPUMD code [50]. At a twist angle of 21.79° (Fig. 4(a)), the energy profile displays a single transition barrier of 1.74 meV/f.u. between two degenerate ferroelectric states with opposite in-plane polarization. At the transition state, a centrosymmetric configuration emerges, where oppositely oriented polar textures in twisted bilayers cancel through lateral mirror symmetry. The corresponding phonon dispersion exhibits a pronounced transverse optical (TO) soft mode at the Γ point in the transition state, indicating that the switching is primarily driven by an optical-mode instability. We noticed the unpolarized state in Fig. 3(d) exhibits close energy with the intermediate state in Fig. 4(a), indicating competing transition paths, shown in Fig. S13. At a twist angle of 27.80° (Fig. 4(b)), the energy barrier decreases significantly to 0.18 meV/f.u. The displacement vectors of the two ferroelectric states with upward and downward direction maintain opposite directions, while the polar texture of transition state does not exhibit. This indicates a change in the polarization cancellation mechanism compared to the lower twist angle. Like the 21.79° case, a TO soft mode is observed near the Γ point at the transition state, again revealing an optical-mode-driven phase transition. The polarization textures and phonon dispersions are shown for the three representative configurations. For the highly twisted structure at 46.83° (Fig. 4(c)), the switching path becomes qualitatively different. The energy profile exhibits a nearly flat, wavy form with a maximum barrier of only 0.03 meV/f.u., suggesting an almost barrierless switching process. Six intermediate states are extracted along the path, revealing a continuous evolution of in-plane polarization textures. Unlike the lower twist angles, no evident optical soft modes are observed in the phonon dispersions of the intermediate states. The absence of soft-mode instability suggests that the polarization switching in this case is not driven by conventional lattice instability but rather proceeds through a smooth transformation of local polar textures, which will be further discussed below.

The unit cells of twisted $HfO_2$ with twist angles of 21.79°, 27.80°, and 46.83°, contain 42, 78, and 114 atoms, respectively. The calculated energy barriers are 1.74 meV/f.u., 0.18 meV/f.u., and 0.03 meV/f.u. These values are remarkably low, indicating a superlubric ferroelectric transition enabled by the twisting strategy—particularly for the 27.80° twisted structure, which exhibits a characteristic energy barrier profile commonly seen in ferroelectric materials [51-54]. In the case of the 46.83° twisted structure, the extremely low periodic energy variation circumvents the high energy barrier of ferroelectric switching in bulk $HfO_2$. These results highlight the effectiveness of twist-induced polarization modulation in bilayer $HfO_2$ by polar textures associated with moiré pattern. We note that the energy barriers between ferroelectric states in twisted structures are extremely small compared to room-temperature thermal kinetic energy of about 0.0243 eV/atom. But unlike conventional ferroelectrics, where on-site polarization switching is easily disrupted by thermal motion above the Curie temperature, sliding ferroelectrics switch via a macroscopic displacement of a few angstroms is more resilient to thermal fluctuations. Besides, a rough estimate from the slope of energy-polarization between the ferroelectric and intermediate unpolarized states suggests electric fields of $1.4\times10^5$ V/cm, $1.8\times10^4$ V/cm, and $8.8\times10^2$ V/cm for 21.79°, 27.80°, and 46.83° twisting configurations to drive the ferroelectric transition. If 10 μm-sized device is considered, the corresponding voltage for the 21.79° configuration is approximately 1.4 V, which is experimentally accessible.

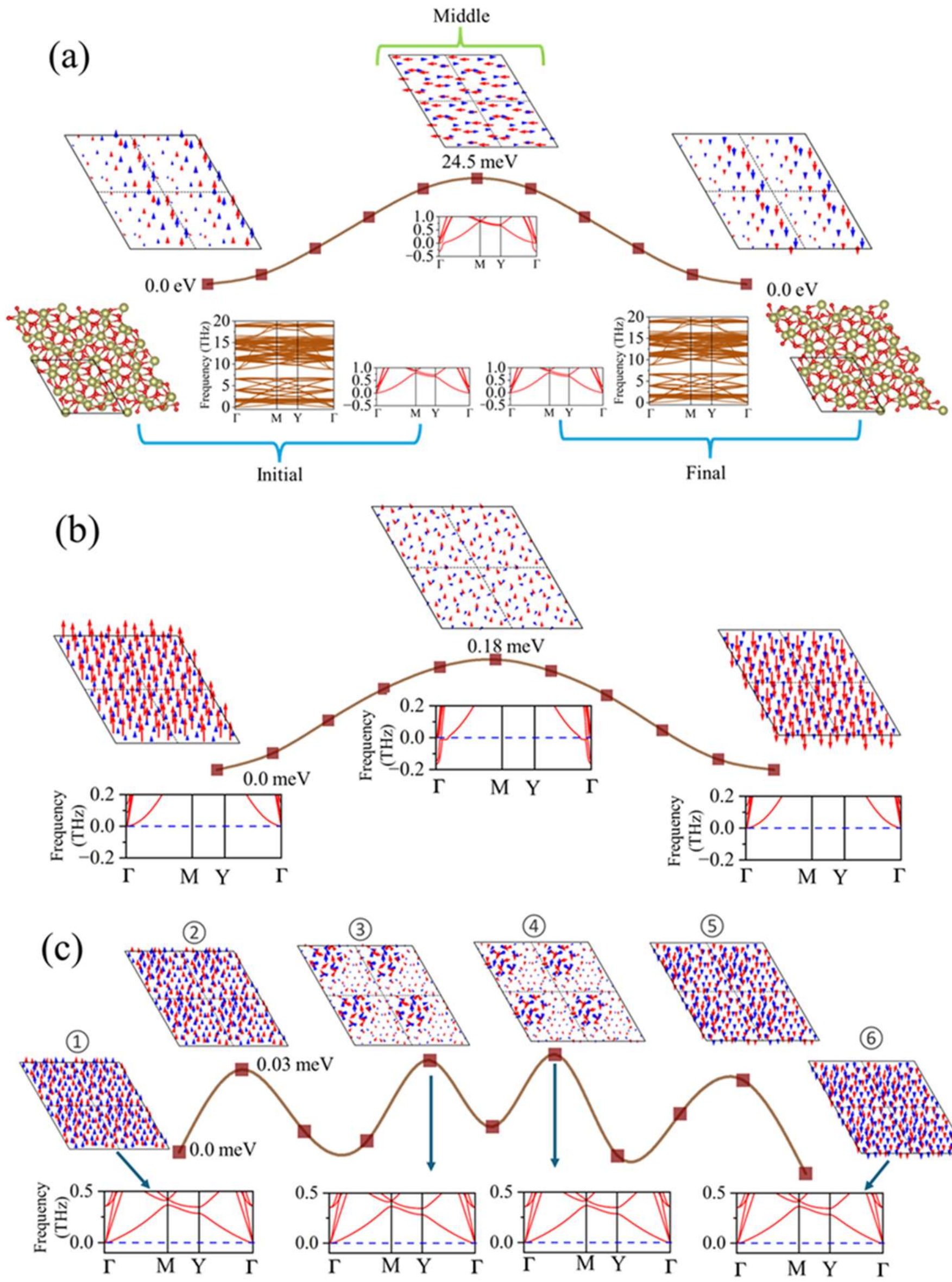

**Fig. 4. Switching pathways in twisted bilayer HfO₂.** Ferroelectric switching pathways of twisted bilayer $HfO_2$ with twist angles of (**a**) 21.79°, (**b**) 27.80°, and (**c**) 46.83°, obtained from NEB calculations. Each panel shows the energy profile along the switching path, in-plane polarization patterns (red and blue arrows) at representative images, and the corresponding phonon dispersions with enlarged views of low-frequency branches.

**Phonon linewidth**

Traditional descriptions of ferroelectric phase transitions often rely on the softening of optical phonon modes, while neglecting the role of phonon-phonon anharmonic interactions.

To explore the origin of the twist-angle-dependent switching barrier, particularly the ultralow barrier at 46.83°, we calculated the phonon linewidths of twisted bilayer $HfO_2$ at 300 K using normal mode analysis based on the spectral energy density (SED) method [55-57]. Detained information of this method is seen in Methods. Molecular dynamics (MD) simulations were performed using the GPUMD package [50] with NEP potentials to extract phonon linewidth *via* SED analysis using Dynaphopy code [58]. A 5×5×1 supercell of each twisted structure was first relaxed in the NPT ensemble for 200 ps. Atomic trajectories were subsequently extracted from a 100 *ps* NVT ensemble simulation with a 1 *fs* timestep.

As shown in Fig. 5(a), increasing the twist angle from 21.79° to 46.83° leads to a clear enhancement in phonon linewidths across the entire frequency range, especially in the low-frequency region. This indicates stronger anharmonic phonon-phonon interactions at larger twist angles. Fig. 5(b) highlights the enlarged view of the transverse optical (TO) branch near the Γ–M path. The linewidth increases from 0.014 $ps^{-1}$ at 21.79° and 0.017 $ps^{-1}$ at 27.80°, to 0.184 $ps^{-1}$ at 46.83°, by an order of magnitude larger. Such strong anharmonicity can drive the softening of optical modes more effectively toward zero energy, thereby promoting the occurrence of the phase transition.

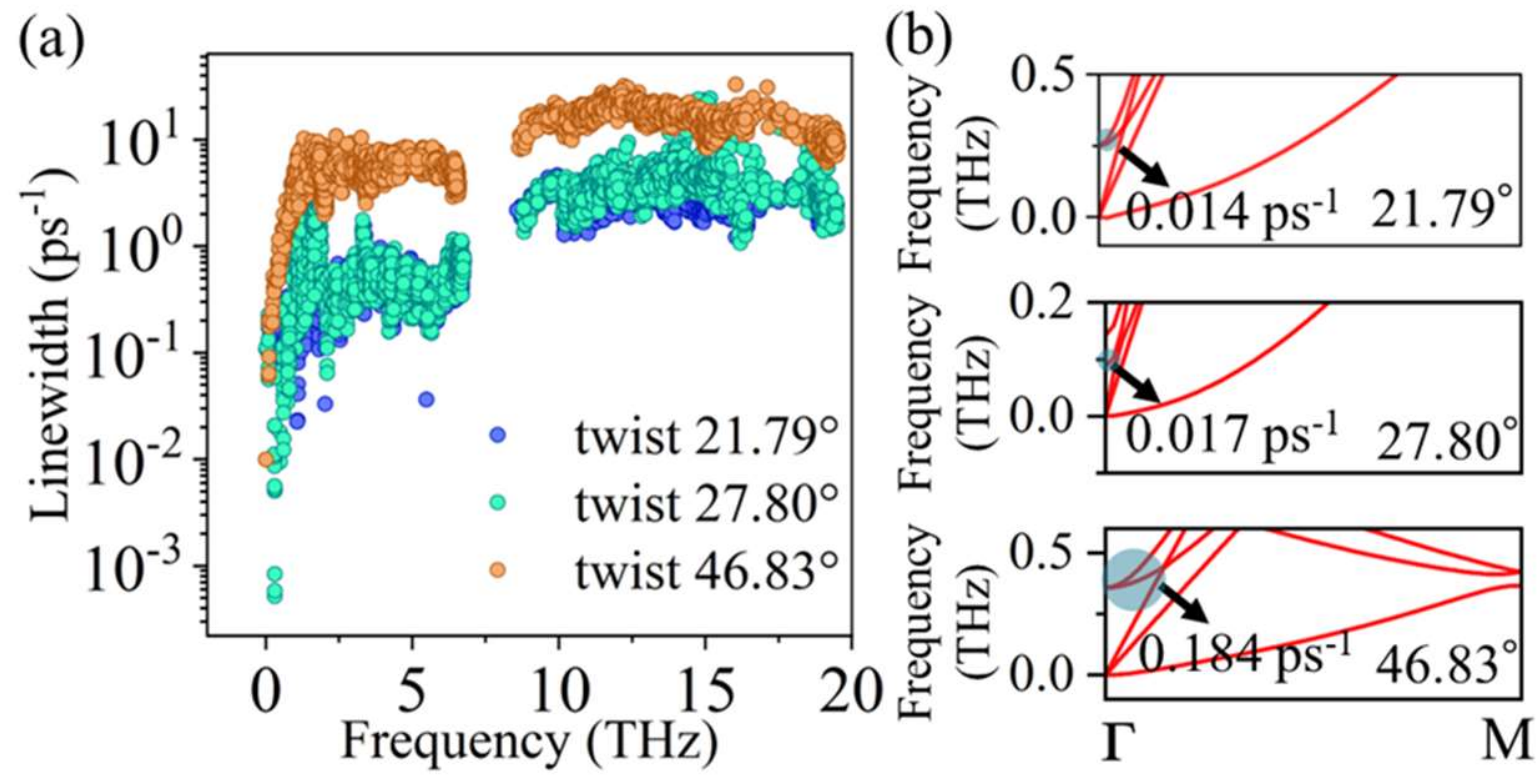

**Fig. 5. Anharmonicity of twisted bilayer $HfO_2$.** (**a**) Phonon linewidths of twisted bilayer $HfO_2$ with twist angles of 21.79°, 27.80°, and 46.83° at 300 K. (**b**) Enlarged phonon dispersions along the Γ→M path for the three twist angles. The blue shaded regions indicate the linewidths of the transverse optical (TO) branch.

**Discussion**

In this work, We demonstrated a machine-learning-assisted approach to reveal the ferroelectric switching behavior in bilayer $HfO_2$ under sliding and twisting, leveraging MLIP for both efficient phonon dispersion calculations via force constants and phonon linewidth extraction via SED analysis from MD simulation. For the first time, we identified a moiré-induced optical soft mode as the driving force of the twist-enabled ferroelectric phase transition with superlubric-like energy barrier. Furthermore, the pronounced enhancement of anharmonicity in twisted configurations was found to play a crucial role in lowering the energy barrier, suggesting a fundamental mechanism behind the facilitated switching in moiré-engineered 2D ferroelectrics. These findings provide new insights into designing ultralow-energy ferroelectric devices through topological and anharmonic control.

## Methods

### First-principles calculation

The structural relaxation of monolayer and bilayer $HfO_2$ with and without twisting is conducted using density functional theory (DFT) with the projector augmented wave (PAW) method,[59] as implemented in the Vienna Ab initio Simulation Package (VASP).[60,61] To describe electronic exchange and correlation effects, we adopt the Perdew-Burke-Ernzerhof (PBE) functional within the generalized gradient approximation (GGA).[62] The D3 dispersion correction, as proposed by Grimme, is incorporated to account for interlayer van der Waals interactions.[63] The energy cutoff of 550 eV is used for all the structures. The Brillouin zone is sampled using Monkhorst-Pack *k*-point grids of 11×11×1 and 7×7×1 is used for the untwisted and twisted structures, respectively. Structural optimization follows convergence thresholds of $10^{-6}$ eV/Å for atomic forces and $10^{-8}$ eV for total energy. Additionally, a vacuum layer of 25 Å is introduced along the out-of-plane direction. The spontaneous polarization is calculated based on the modern theory of Berry phase.[35,36] The ferroelectric switching pathway is determined through calculations based on the climbing-image nudged elastic band (CI-NEB) method.[41]

### Strain gradient application

Considering the common presence of strain gradients in twisted systems, we started by introducing two different strain formulations to explore the vortex pattern in twisted $HfO_2$, a strain gradient in equation S1, as shown below,

$$x' = x[1 + \varepsilon_0 \times (1 + \alpha x)] \quad (1)$$

Where $x$ and $x'$ represent the original and strained atomic coordinates in the $x$-direction ($a$-direction), respectively. $\varepsilon_0$ is the uniform strain applied to the system, while $\alpha$ is a strain gradient parameter that introduces a position-dependent strain variation. The term $\alpha x$ represents the spatial dependence of strain, making the deformation vary with position. Here, we set $\varepsilon_0$ as 3% and $\alpha$ as 0.5 as a case study.

**Machine-learning potential**

We employed the Neuroevolution Potential (NEP)[49] to model interatomic interactions in bilayer $HfO_2$, both with and without twisting. To generate training structures for non-twisted bilayer $HfO_2$, we firstly applied the Monte Carlo displacement method implemented in the Hiphive package,[34] obtaining 60 configurations. The supercell was set to 5×5×1, containing 150 atoms per structure. For twisted bilayer $HfO_2$, smaller supercells were used because the moiré lattices are constructed from the supercells of untwisted structure. Specifically, 2×2×1 supercells were generated for the 21.79° and 27.80° twisted structures, comprising 168 and 302 atoms, respectively. For the 46.83° twisted $HfO_2$, the primitive unit cell with 114 atoms was utilized as the training structure. DFT calculations were then performed on these structures to obtain reference energy and interatomic force data. The $k$-point mesh was set to 3×3×1 for non-twisted, 21.79°, and 46.83° twisted bilayer $HfO_2$, while a 2×2×1 $k$-point grid was applied for the 27.80° twisted structure. During NEP training, radial and angular cutoffs were set to 8 Å and 5 Å, respectively, while all other parameters remained at their default values. The training process spanned $10^6$ steps. The comparisons between DFT and NEP predictions for energy, atomic forces, and virial stress are shown in the Fig. S9, demonstrating strong agreement.

**Spectral energy density calculations**

The phonon normal mode coordinate $Q(q,s,t)$ is constructed by projecting atomic velocities onto the phonon eigenvectors:

$$Q(q,s,t) = \sum_{jn} \sqrt{\frac{m_j}{N}} e_j^*(q,s) v_{jn} \exp(-2\pi i q \bullet r_l) \ , \qquad (2)$$

where $q$ is the wave vector, $s$ denotes the phonon branch index, $m_j$ the mass of atom $j$, $N$ is the number of unit cells, $v_{jn}$ is the velocity of atom $j$ in the $n$-$th$ unit cell at time $t$, $r_l$ is the position of that unit cell, and $e_j^*$ is the complex conjugate of the phonon eigenvector. The SED function $\phi(q,s,f)$ is then obtained by performing a Fourier transform of the time-dependent mode amplitude:

$$\phi(q,s,f)=\left|\int Q(q,s,t)e^{-2\pi ift}dt\right|^2, \tag{3}$$

The resulting spectrum shows Lorentzian peaks centered at the phonon frequencies. Each peak is fitted using the following expression:

$$\phi(q,s,f)=\frac{I(q,s)}{[f-f_0(q,s)]^2+\Gamma^2(q,s)}, \tag{4}$$

where $f_0$ is the peak position and $\Gamma(q, s)$ is the half-width at half-maximum (HWHM). The phonon linewidth is given directly by $2\Gamma(q, s)$.

## Data availability

The datasets generated in this work are available upon request.

## Acknowledgements

Y. S. acknowledges the project funded by the Science and Technology Commission of Shanghai Municipality (No. 24CL2901702). We acknowledge the Supercomputer Center, Institute for Solid State Physics, the University of Tokyo (Project number 2024-Cb-0042).

## Competing interests

The authors declare that they have no competing interests.

**Supplementary Information**

**Section 1.** Electronic structure of monolayer $HfO_2$

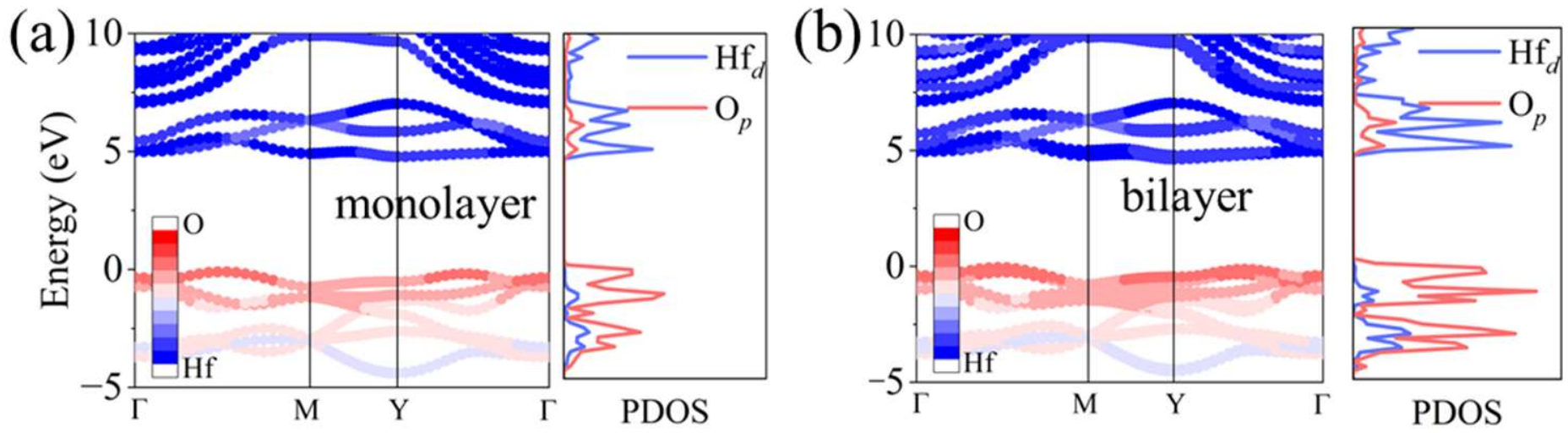

Fig. S1. Projected band structures and density of states of (a) monolayer $HfO_2$ and (b) bilayer $HfO_2$ (AA').

**Section2.** Electronic structures of polarized bilayer $HfO_2$

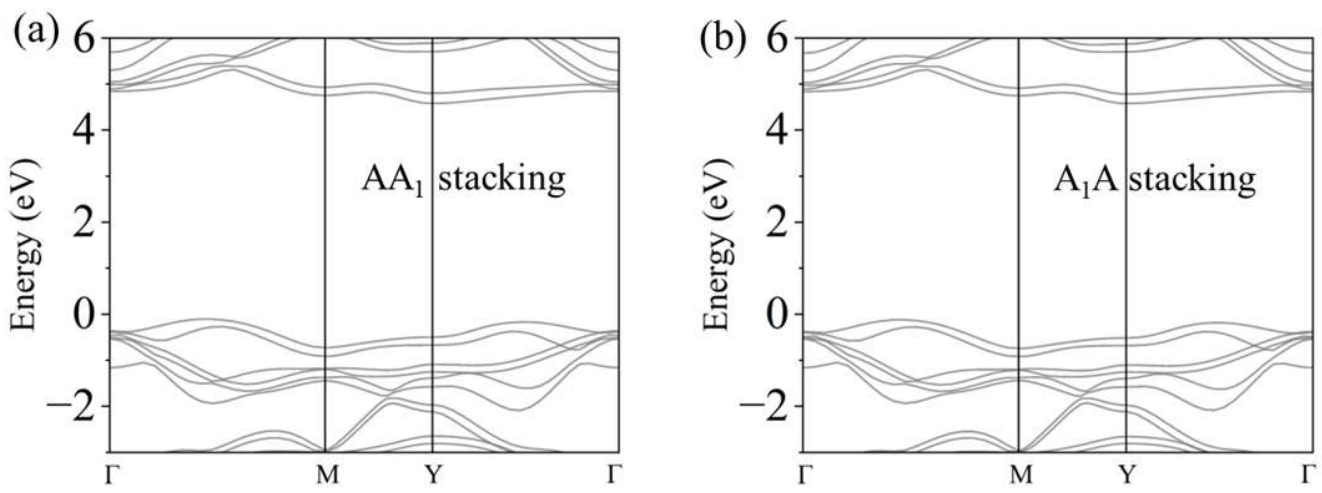

Fig. S2. Electronic structures of the bilayer $HfO_2$ under $AA_1$ and $A_1A$ stacking pattern.

**Section 3.** Soft optical mode in paraelectric phase

The phonon dispersion of the intermediate paraelectric state exhibits the soft optical mode, as shown in Fig. S3(a). We calculated the eigenvectors of the paraelectric phase and then visualized the atomic displacements of the intermediate state [1], as shown in Fig. S3(b) and (c). The related vibration of soft optical mode induces interlayer sliding and thus polarization. These displacements in the intermediate structure ultimately give rise to the ferroelectric $AA_1$ stacking.

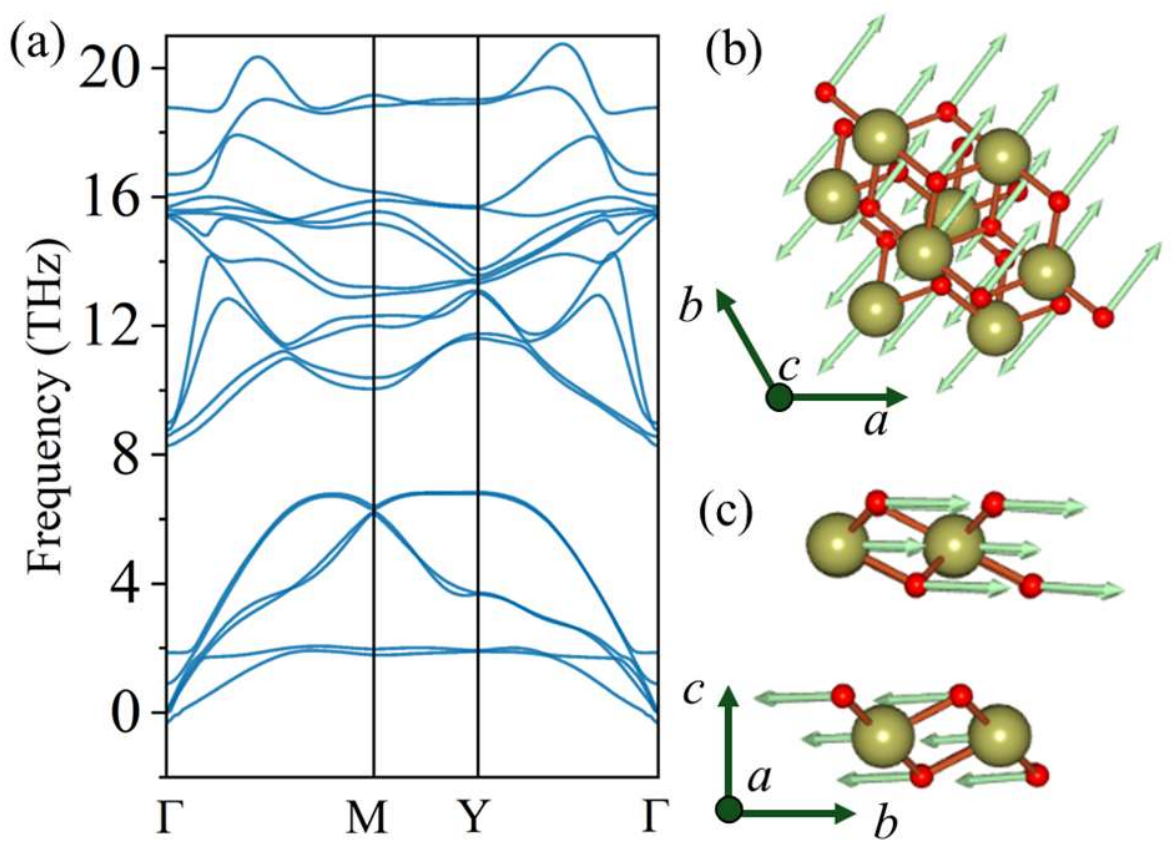

Fig. S3. (a) Phonon dispersion of the intermediate state between two ferroelectric configurations. (b, c) Top and side views of the vibrational direction (green arrow) of soft optical mode at Γ point as obtained from its eigenvectors.

**Section 4.** Geometric structures along the sliding transition path

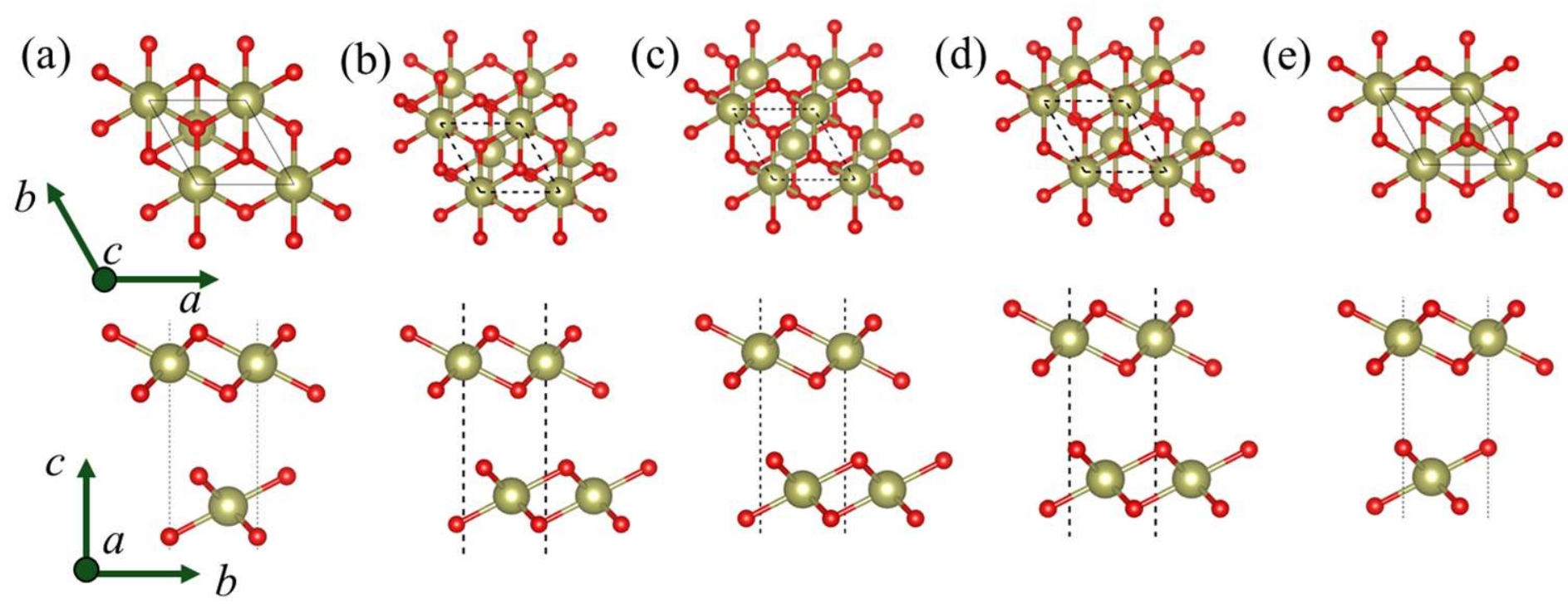

Fig. S4. Top and side views of (a) the $AA_1$ stacking, (b-d) the intermediate states along the transition path (sliding index from 4 to 8), and (c) the $A_1A$ stacking of $HfO_2$.

**Section 5.** Geometric structures of the twisted $HfO_2$

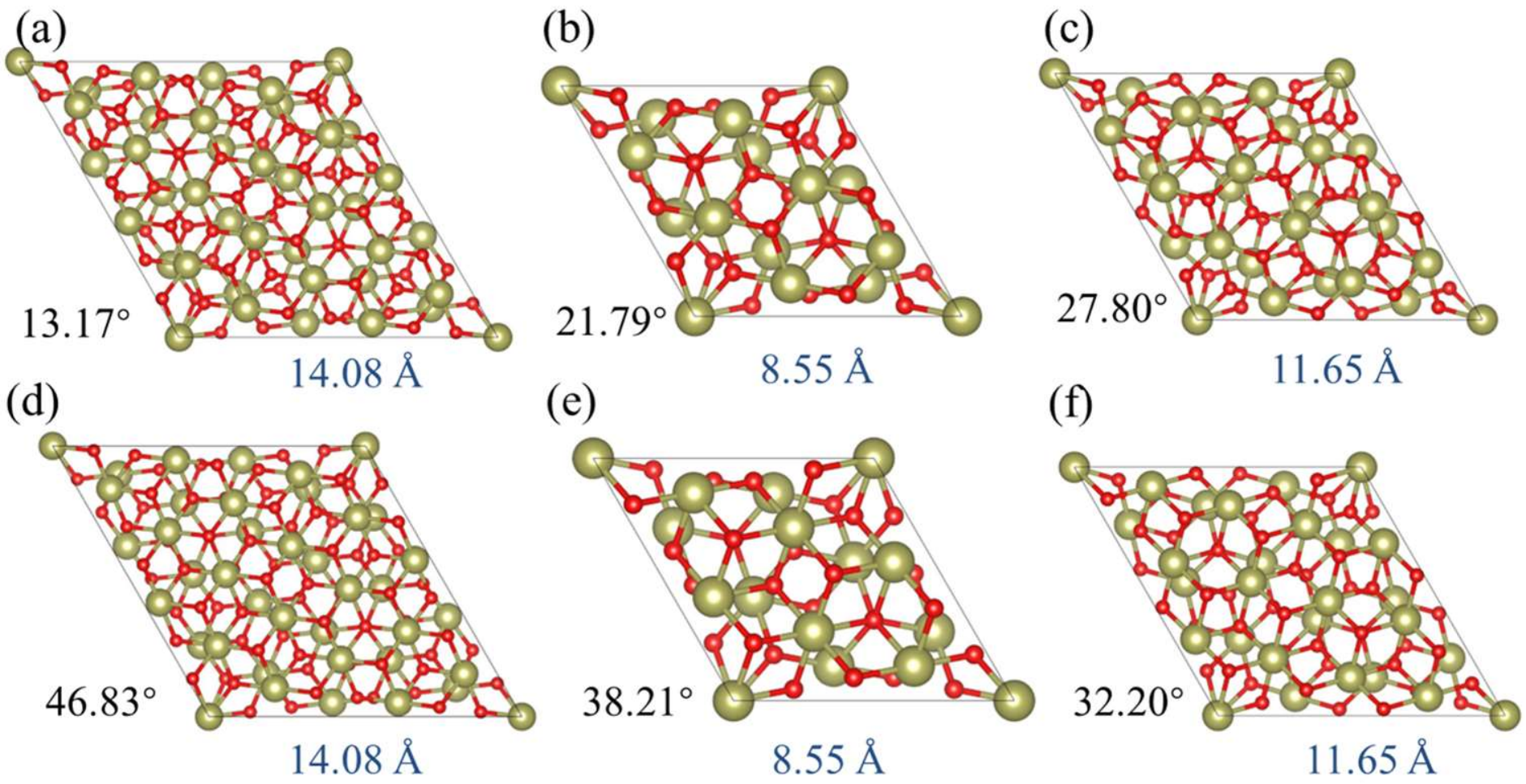

Fig. S5. Top views of the geometric structures of twisted $HfO_2$ with twisting angles of 13.17°, 21.79°, 27.80°, 32.20°, 38.21°, and 46.83°

**Section 6.** Vortex of O atom along the out-of-plane directions

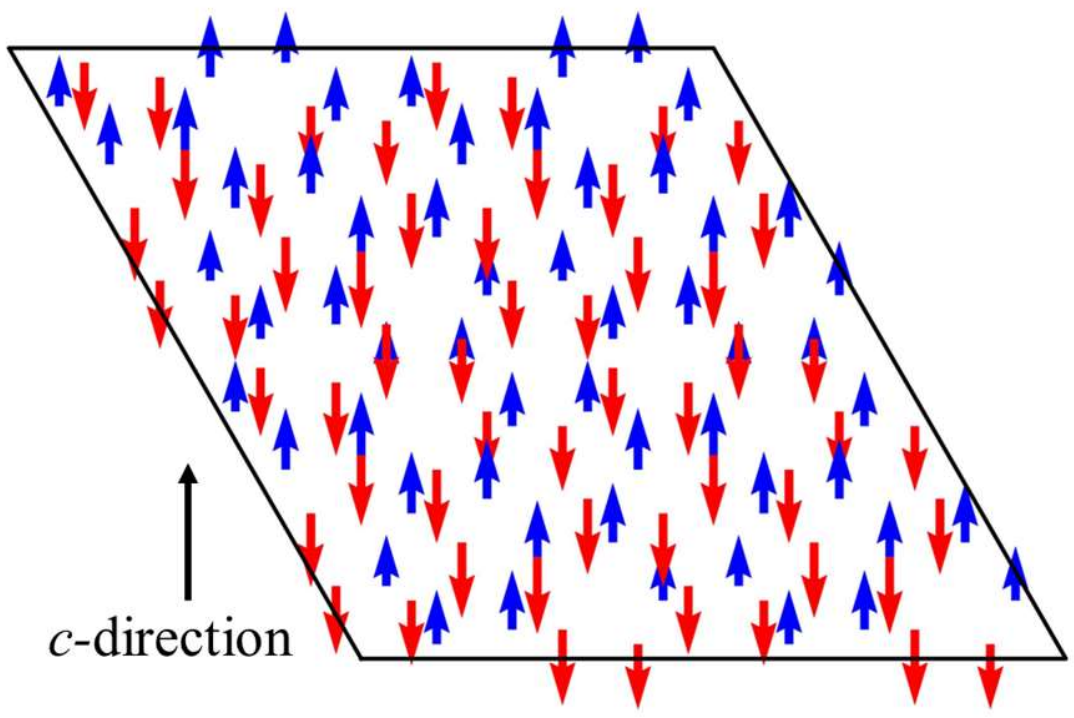

Fig. S6. The displacement of O atoms in twisted 21.79° $HfO_2$ along the out-of-plane direction (*c*-direction). The displacement is magnified by a factor of ten for improved clarity.

**Section 7.** Phonon dispersions with structures after no optimization

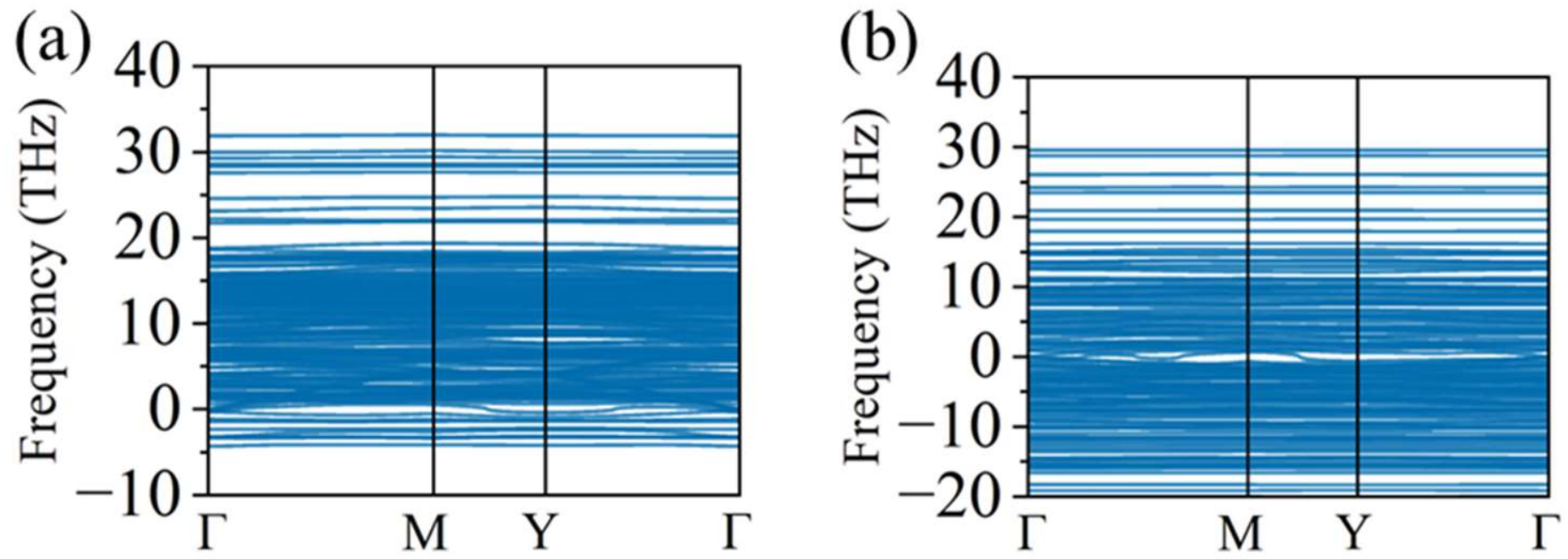

Fig. S7. The phonon dispersions of twisted $HfO_2$ with a twisting angle of 21.79° under (a) non-linear strain gradient and (b) linear strain gradient without optimization.

**Section 8.** Geometric structures under inversed strain

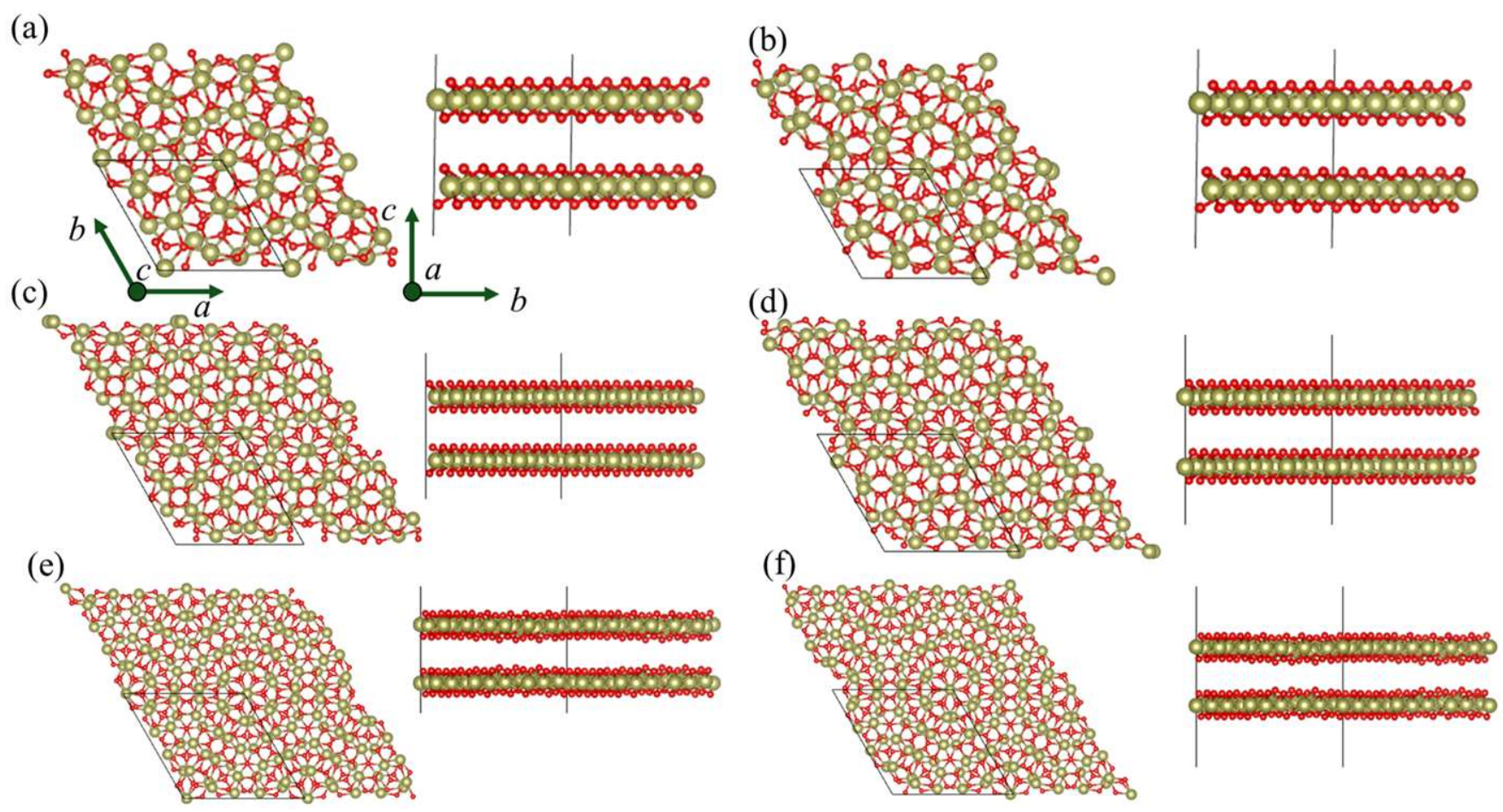

Fig. S8. Top and side views of optimized twisted $HfO_2$ structures (2×2×1 for clearer view) under strain along the +*a* and -*a* directions: (a, b) 21.79° twist angle, (c, d) 27.80° twist angle, and (e, f) 46.83° twist angle.

**Section 9.** NEP training tests

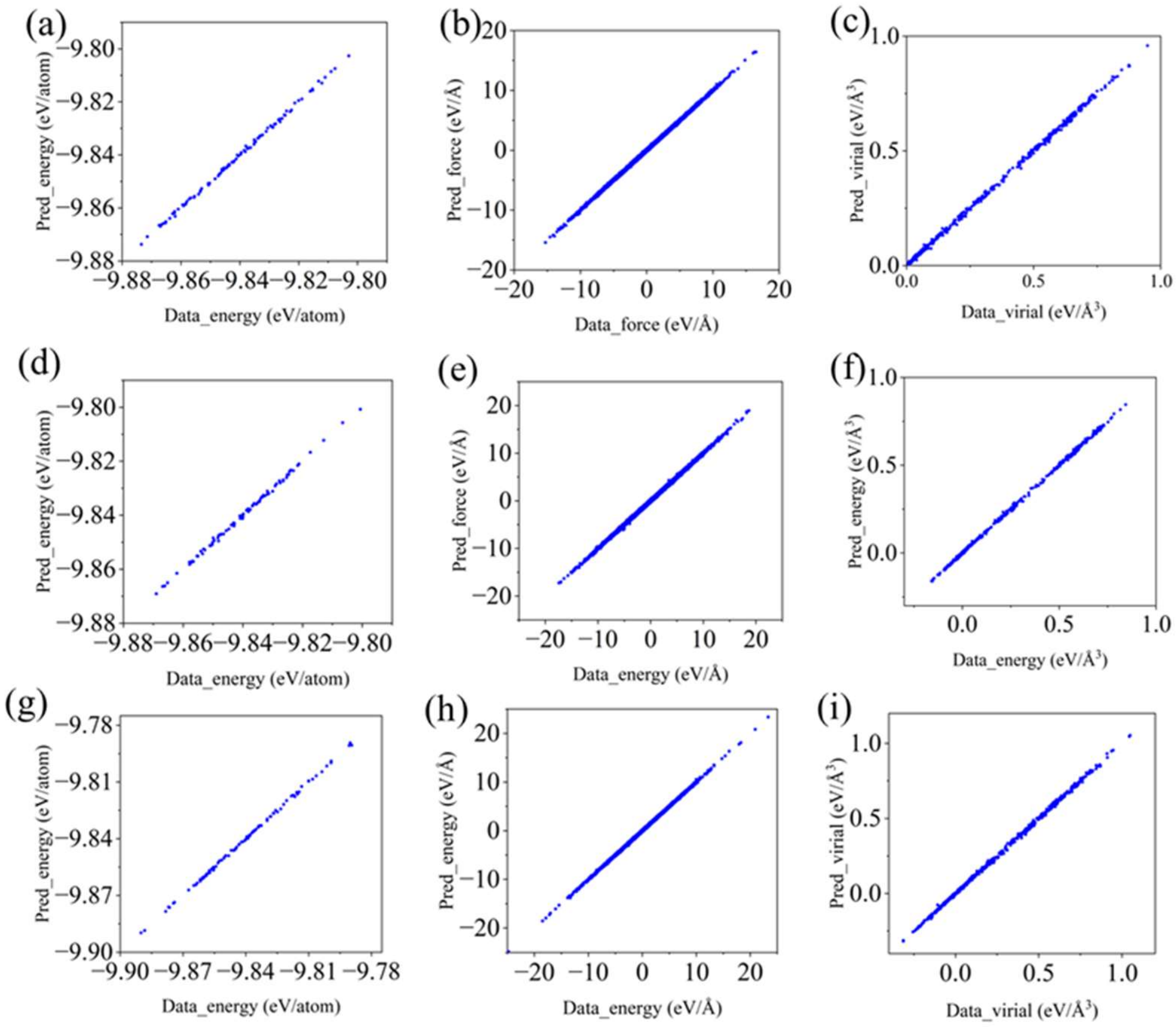

Fig. S9. The energy, force and virial difference between the training data from DFT calculations and the predicted data of (a) twisted $HfO_2$ with a twisting angle of (a-c) 21.79°, (d-f) 27.89° and (g-i) 46.83°.

**Section 10.** Phonon dispersions of twisted structures

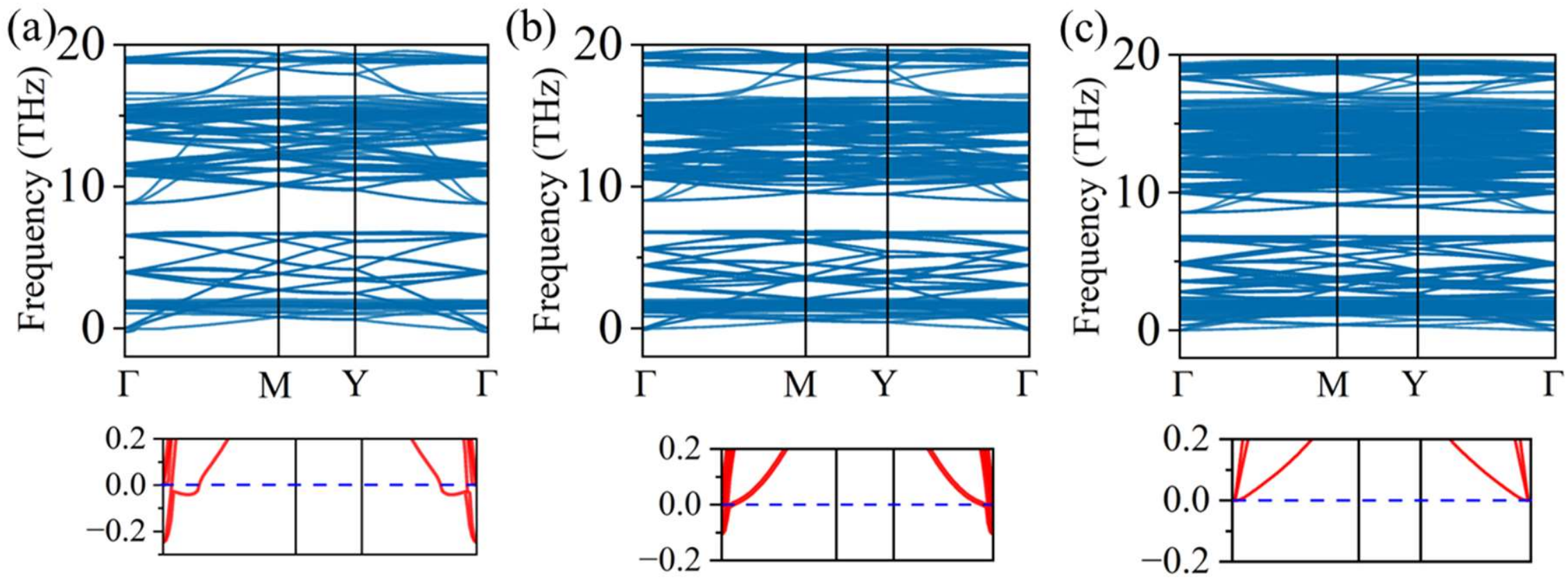

Fig. S10. Phonon dispersions of twisted bilayer $HfO_2$ with a twisting angle of (a) 21.79°, (b) 27.89° and (c) 46.83° based on NEP potential. The enlarged vibrational frequency shown below each phonon dispersion plot provides a clearer view of the low-frequency regions.

**Section 11.** Phonon dispersions of twisted $HfO_2$ after strain

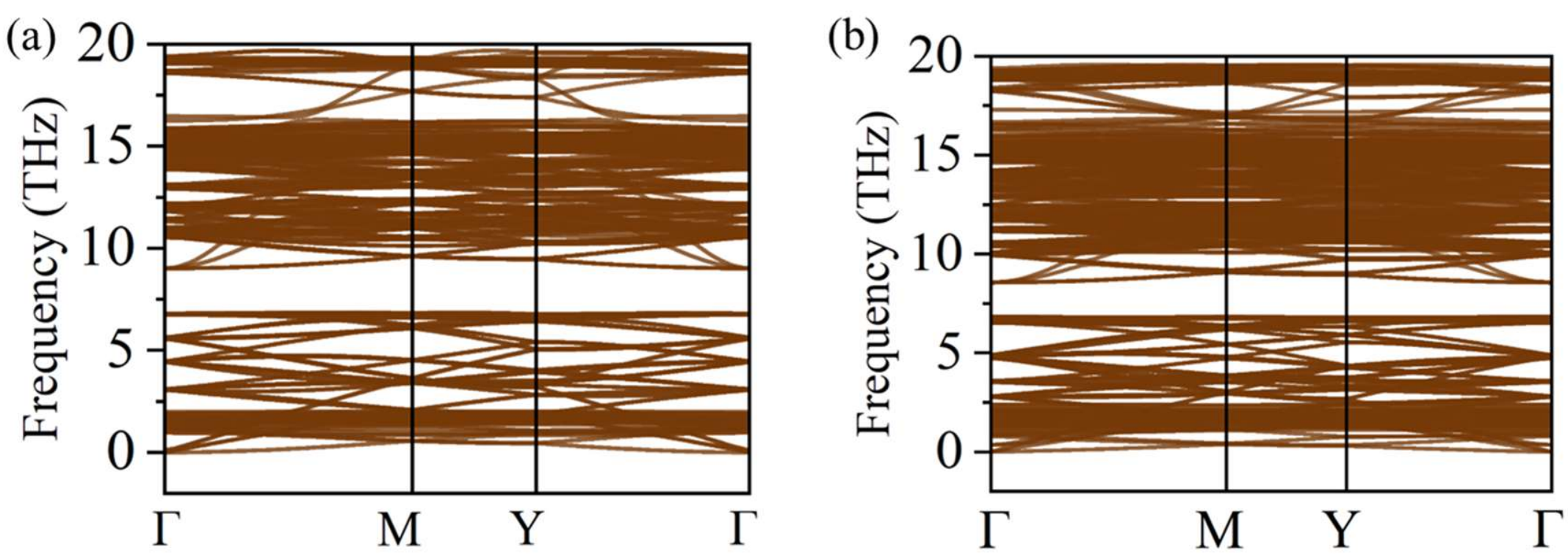

Fig. S11. Phonon dispersions of $HfO_2$ with twist angles of (a) 27.80° and (b) 46.83° after strain perturbation.

**Section 12.** Optimized structures after linear strain applied and phonon dispersions

In equation S1, the term $\alpha\varepsilon_0$ is set to 0.5, representing the slope of the strain gradient, which is much larger than the value of 0.015 used in our case study, where $\varepsilon_0$ is 3% and $\alpha$ is 0.5.

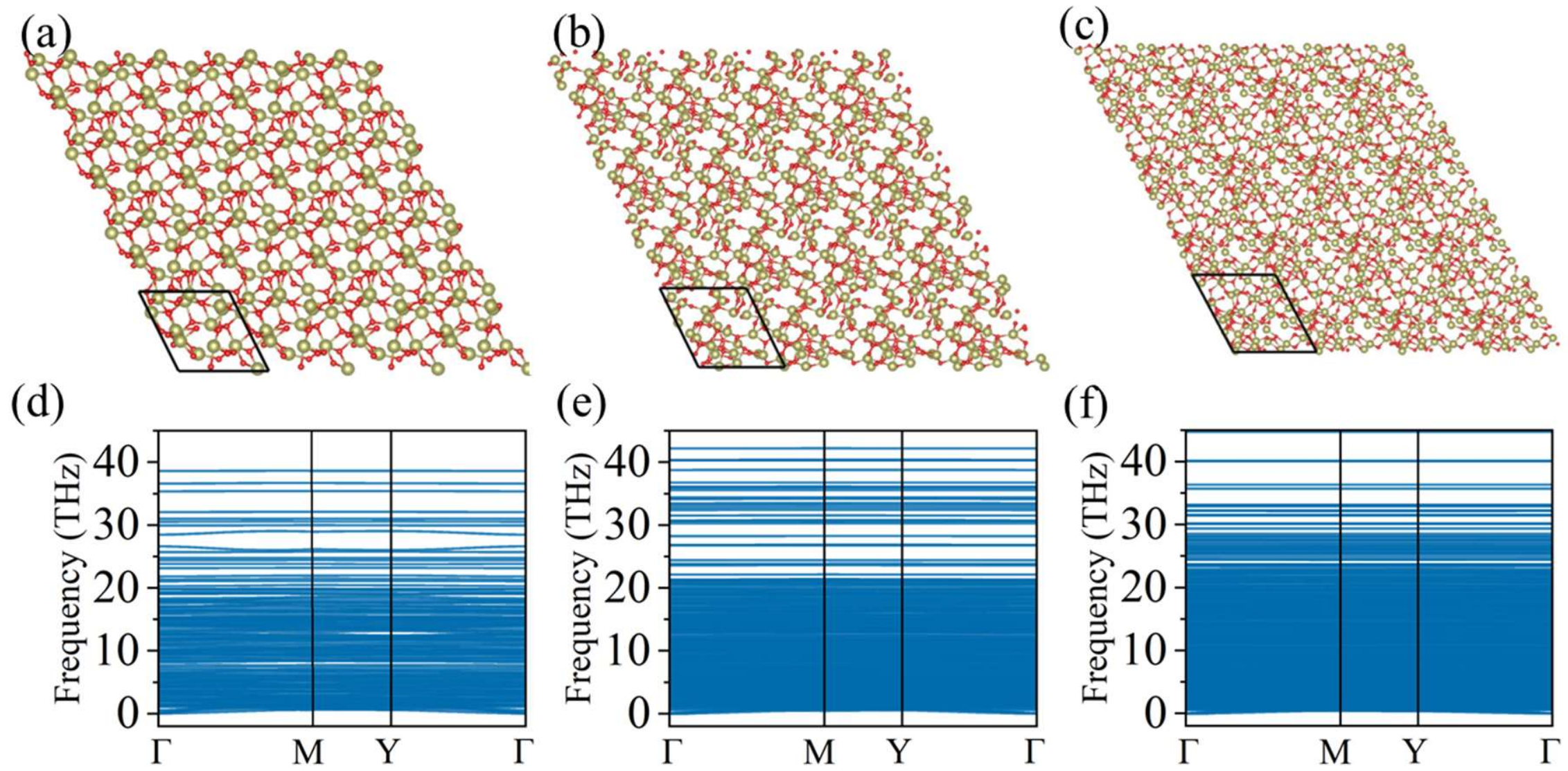

Fig. S12**.** The optimized twisted $HfO_2$ with a twisting angle of (a) 21.79°, (b) 27.89° and (c) 46.83° based on NEP potential after linear strain gradient is applied. (d-f) The corresponding phonon dispersion for each twisted structure.

**Section 13.** Ferroelectric switching path of bilayer $HfO_2$ at a twisting angle of 21.79°

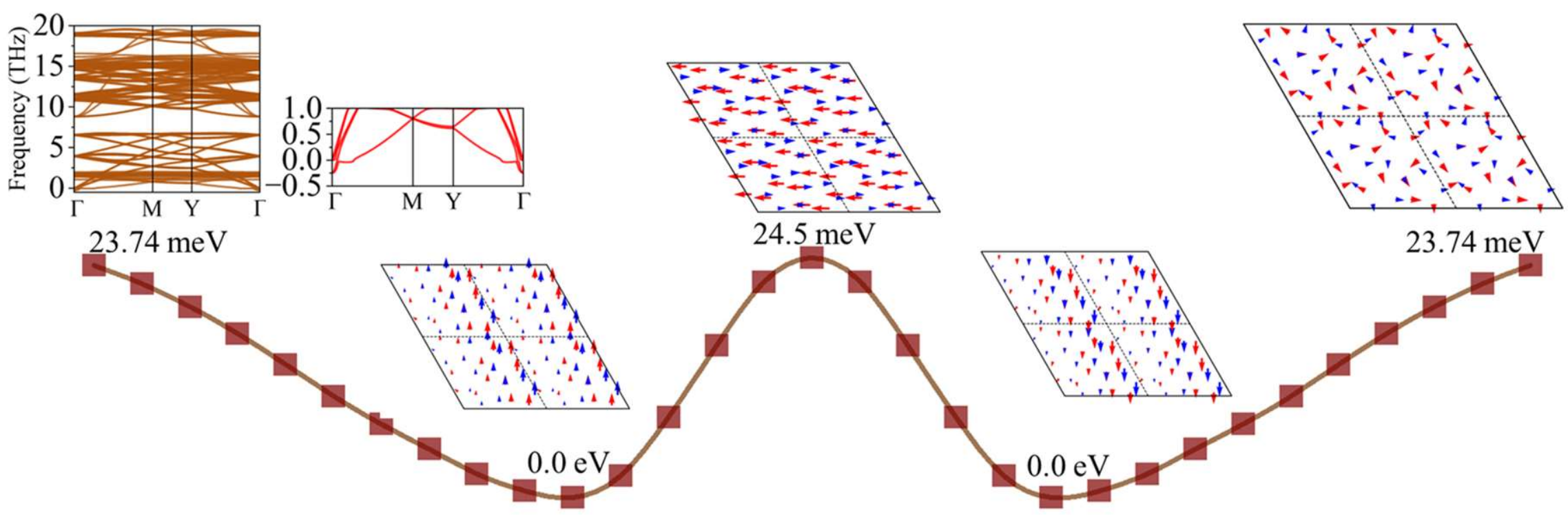

Fig. S13. Ferroelectric switching pathways of twisted bilayer $HfO_2$ with twist angles of 21.79° having different unpolarized intermediate states.